\documentstyle[12pt]{article}
\begin{document}

\begin{titlepage}

\begin{center}
\hfill La Plata-Th 96/14\\
\vspace{0.3in}
{\LARGE\bf On the Path Integral Representation for Spin Systems}
\\[.4in]
{\Large Daniel C. Cabra$^a$, Ariel Dobry$^b$, Andr\'es Greco$^b$
and Gerardo L. Rossini$^a$}

\bigskip {$^a$\it Departamento de F\'\i sica,  \\
Universidad Nacional de La Plata,\\
C.C. 67 (1900) La Plata, Argentina.}

\bigskip {$^b$ \it Departamento de F\'{\i}sica,
Facultad de Ciencias Exactas, Ingenier\'{\i}a y
Agrimensura
and IFIR (CONICET-UNR). Av. Pellegrini 250,
2000 Rosario, Argentina.}
\bigskip\\
\end{center}
\bigskip

\newpage

\centerline{\bf ABSTRACT}
\begin{quotation}
We propose a classical constrained Hamiltonian theory
for the spin. After the Dirac treatment we show that due to the existence
of second class constraints the Dirac brackets of the proposed theory
represent the commutation relations for the spin.
We show that the corresponding partition function, obtained via the
Fadeev-Senjanovic procedure,
coincides with the one obtained using coherent states.
We also evaluate this partition function for the case of a single spin
in a magnetic field.
\end{quotation}
\end{titlepage}
\clearpage

\newpage

\section{Introduction}

The Heisenberg Model has been used for many years as a canonical
model of magnetism in solids. After the discovery of High Tc
superconductivity a renewed interest in this model emerged. This is
principally due to the connection between the antiferromagnetic Heisenberg
model  and other models (Hubbard and $t-J$ models) \cite{hirsch} which are
important to describe the electronic structure of the  High Tc
superconductors.

Although the Heisenberg model is quadratic in
the spin operators it is highly non-trivial.
What turns difficult the solution of the model are the
complicated commutation relations among the spin components. This kind of
commutation relations are indeed characteristic of systems described by
Hubbard operators \cite{hubbard}. In particular the spin algebra ($su(2)$)
is the bosonic subalgebra of that spanned by the Hubbard operators in the
so-called atomic representation of the $t-J$ model, $Osp(2,2)
$ \cite{Wiegmann}.

When one is faced to a system with this commutation rules one usually
implements some kind of decoupling using slave particles \cite{auerbach}. This
new representation is formally exact and solves the above mentioned
difficulty because the basic
variables are conventional bosons and fermions. In particular a path-integral
treatment is  standard. However, in general, no exact calculation can be
carried out thoroughly and some kind of approximation is needed (V.gr. mean
field, large $s$, large $N$, etc.). A natural question which emerges in this framework
(first decoupling and then approximating) is what is an artifact of the
decoupling and what is not (see for example \cite{greco} for a recent
discussion).

For the analysis of this question, a treatment of spin systems without the
introduction of these fictitious slave particles would be of some
importance. In particular, a path integral representation is a suitable
formulation for this analysis. Indeed, such a path integral representation
can be written using the coherent state method (also known as geometric
quantization) \cite{Wiegmann,fradkin}.

Although path integral methods have shown to be very powerful in various
areas, the coherent state method is not a familiar subject in solid
state physics. Moreover, as pointed in the literature \cite
{auerbach,fradkin,shankar}, there are some mathematical
subtleties in the derivation
of this path integral representation that prevent for a complete confidence
on the results.

In this paper we present a constrained Hamiltonian system which after the
Dirac treatment \cite{dirac} reproduce the physics of the spin. Then we
write down its Feynman path integral following the Fadeev-Senjanovic method
for constrained Hamiltonian systems \cite{senjanovic}. Our derivation is
formally exact, independent of the coherent state method
and valid irrespectively of  the value of the
spin s.   Moreover, it has the virtue
of setting the study of spin systems on canonical grounds.
 This is an important point in view of possible further applications of the
formalism for spin 1/2 systems.

Finally we explicitly calculate the free energy of a particle with fixed
spin in an external magnetic field. The result gives an independent check
for the validity of the expression of the partition function for all values
of the spin. Besides, we think that the methods we use are very instructive and could be
useful to develope approximation schemes for the Heisenberg model.

The paper is organized as follows: in section II we present the constrained
Hamiltonian system and we perform the Dirac treatment. In section III we
find the path integral representation for the quantum mechanics and the
partition function.
We calculate the partition function of one spin
in Section IV. Finally  a brief summary of our main results
is given in section V.

\section{The constrained Hamiltonian system}

The starting point for our analysis is the following Hamiltonian

\begin{eqnarray}
H=H(\vec{n})
\end{eqnarray}

and the additional set of primary constraints

\begin{eqnarray}
\Gamma_1 \equiv P_1-s A_1(\vec{n})\approx 0 ,
\end{eqnarray}

\begin{eqnarray}
\Gamma_2 \equiv P_2-s A_2(\vec{n})\approx 0 ,
\end{eqnarray}

\begin{eqnarray}
\Gamma_3 \equiv P_3-s A_3(\vec{n})\approx 0  ,
\end{eqnarray}
where $\vec P$ is the canonical momentum associated to the vector
$\vec n$ and
\begin{eqnarray}
\Gamma_4 \equiv \mid\vec{n}\mid^2-1 \approx 0
\end{eqnarray}

Equations (1-5) define a classical constrained Hamiltonian system in the
Dirac sense \cite{dirac}. Note that the constraints (2-4) are imposed in the
weak sense.

The classical Hamiltonian is a general function of a vector $\vec{n}$. To
simplify the notation we do not include lattice indices in the vector $\vec{n%
}$ but the following analysis is valid in the case where the spins live over
a lattice. The Hamiltonian (1) is independent of the momentum $\vec{P}$
associated to the vector field $\vec{n}$. The corresponding momentum appears
through the constraints (2-4). In these constraints we also have a general
vector function $\vec{A}(\vec{n})$. In (2-4) $s$ is an arbitrary number. It
will become the size of the spin after quantization. Finally, the
constraint (5) indicate the normalization of the vector $\vec{n}$.

According to the Dirac procedure we have to define the following total
Hamiltonian:

\begin{eqnarray}
H_{T}=H(\vec{n})+\sum_i \Gamma_i \lambda_i
\end{eqnarray}
where the Lagrangian multipliers $\lambda_i$ are general functions of the
coordinates and momenta.

The consistency conditions of the theory require the preservation in time of
the constraints (2-5)

\begin{eqnarray}
\dot{\Gamma_i} = \{\Gamma_i,H_{T}\}_{PB} \approx 0
\end{eqnarray}

In (7) $\{,\}_{PB}$ means the conventional Poisson Brackets. It is
straightforward to show that the system (7) determines uniquely the Lagrange
multipliers (the explicit form of these multipliers is not necessary for
the rest of the paper). Therefore the
constraints are preserved in time and secondary constraints are not
generated. Moreover our theory is completely determined by the first class
Hamiltonian (6) and the second class constraints (2-5) (there is no hidden
gauge invariance in our system).

Due to the second class character of the constraints, the next step is the
calculation of the Dirac brackets (DB). For any two classical quantities $A
$, $B$, the Dirac bracket is defined by

\begin{eqnarray}
\{A,B\}_{DB} = \{A,B\}_{PB} - \{A,\Gamma_i\}_{PB} \Delta_{ij}
\{\Gamma_j,B\}_{PB},
\end{eqnarray}
where

\begin{eqnarray}
\Delta_{ij} \{\Gamma_j,\Gamma_k\}_{PB} =\delta_{ik}.
\end{eqnarray}

Using the PB, it can be shown that the antisymmetric matrix $\Delta^{-1}_{ij}
$ has elements

\begin{eqnarray}
\Delta^{-1}_{12}&=& s[\frac{\partial A_2}{\partial n_1}- \frac{\partial A_1}{%
\partial n_2}],  \nonumber \\
\Delta^{-1}_{13}&=& s[\frac{\partial A_3}{\partial n_1}- \frac{\partial A_1}{%
\partial n_3}]  \nonumber \\
\Delta^{-1}_{14}&=&-2 n_1  \nonumber \\
\Delta^{-1}_{23}&=& s[\frac{\partial A_3}{\partial n_2}- \frac{\partial A_2}{%
\partial n_3}]  \nonumber \\
\Delta^{-1}_{24}&=&-2 n_2  \nonumber \\
\Delta^{-1}_{34}&=&-2 n_3
\end{eqnarray}

In summary, we have a theory which is defined by a first class Hamiltonian
(6) without arbitrary coefficients and the set of second class constraints
(2)-(5).

At this point the constraints are imposed as strong equations between
coordinates and momenta. Also the PB are replaced by the DB.

\section{Quantization and Statistical Mechanics}

The canonical quantization can be now performed in the standard way. The
Hamiltonian (6) have to be considered as the Hamiltonian operator. The
constraints (2-5) become strong equations between operators and the
commutation relations are obtained from the DB. The commutation relations
between two operators ${\hat{O}}_1$ and ${\hat{O}}_2$ are given by:

\begin{eqnarray}
[\hat{O}_1,\hat{O}_2]=\imath \hbar \{O_1,O_2\}_{DB}.
\end{eqnarray}

The vector field $\vec{A}(\vec{n})$ is not specified yet. If we choose it as:

\begin{eqnarray}
\nabla_{\vec{n}} \wedge \vec{A}(\vec{n})= \vec{n},
\label{rot-a}
\end{eqnarray}
the DB between the components of $\vec{n}$ reads:

\begin{eqnarray}
\{n_i,n_j\}_{DB}=\frac{n_k}s.
\end{eqnarray}
In the last expression ijk are the integers 123 and its cyclic permutations.

If we now identify  $s \hat{n}_i\rightarrow \hat{S}_i$, we obtain:

\begin{eqnarray}
[\hat{S}_i,\hat{S}_j]=\imath \hbar \hat{S}_k
\end{eqnarray}
which  corresponds  to the commutation relations of the
components of the spin operator. Then, we have proposed a
classical constrained theory which after the Dirac procedure reproduce the
commutation relations for the spin.

Now, we are able to quantize the system using Feyman path integral. We shall use
the method developed for constrained systems in \cite{senjanovic}. In this
method the probability amplitude of the system which was at $\vec{n_0}$
at $t=0$ and  will
be at $\vec{n_1}$ at time $T$ can be written as:

\begin{eqnarray}
<\vec{n_0}\mid\vec{n_1}>=\int d\vec{n} d\vec{P} (\prod_i \delta(\Gamma_i))
(det \Delta^{-1})^{1/2} exp(\frac{\imath}{\hbar} {\int_{0}}^T dt [\vec{P}
\dot{\vec{n}} -H_T])
\end{eqnarray}

Using the fact that $det(\Delta)$ is a constant independent of the fields
(see (10)) and integrating over $\vec{P}$, we obtain

\begin{eqnarray}
<\vec{n_0}\mid\vec{n_1}>=\int d\vec{n} \delta(\mid\vec{n}\mid^2-1) exp(\frac{%
\imath}{\hbar} {\int_{0}}^T dt [s\vec{A}(\vec{n}) \dot{\vec{n}} -H(\vec{n})]).
\end{eqnarray}

To obtain the partition function, we must carry out the integration over
all periodic paths, make the change $it=\tau$ and identify $iT/\hbar$ 
with the inverse of the temperature $\beta$  \cite{bernard}. Thus we
have

\begin{eqnarray}
Z=\int_{periodic} d\vec{n} \delta(\mid \vec{n} \mid^2-1) exp(-\frac{1}{\hbar} {\int_{0}}%
^{\beta\hbar} d\tau [-s \vec{A}(\vec{n}) \frac{d\vec{n}}{d\tau} +H(\vec{n})]).
\label{zeta}
\end{eqnarray}
Equations (16) and (17), with $\vec A$ given by (12), are exactly the
expressions for the path
integral for the spin system obtained via coherent states \cite{fradkin}.

Before going to the next section it is useful to make a few remarks
about the reason for choosing the set of constrains (2)-(5). Our
principal aim was to find a classical system containing the nature of
the spin. To get a chance to find that, the classical
system must contain constraints, and these constraints must be second
class. 
The set of constraints (2)-(5) are {\it arbitrary} functions of the coordinates 
and momenta and our previous results showed 
that after imposing condition (12)  
the quantized theory corresponds
to the spin system. 
On the other hand, the   classical constrained system we have introduced 
describe a system of massless particles living on 
unitary spheres. Each particle feels a magnetic field represented
by the vector potential $\vec{A}$. This vector potential corresponds
to the one of a monopole placed at the center of the sphere as is 
stressed by condition (12). This mechanical analogy was previously
suggested from the path integral expression obtained via coherent
states \cite{fradkin} and has indeed inspired the election of our set of 
constraints.         
It is important to point out that once 
the constrained Hamiltonian 
system was defined we were able
to show that the path integral representation is valid for any
parameter $s$. This is a point  not allways clear in the literature.

\section{An example}

Our starting point is eq.(\ref{zeta}), with the Hamiltonian
\begin{eqnarray}
H(\vec{n})=-s\vec{n}.\vec{B}. \label{5.1}
\end{eqnarray}
The first term in the action of eq.(\ref{zeta}) can be written, with the
aid of eq.(\ref{rot-a}) and Stoke's theorem, as
\begin{eqnarray}
\int_0^\beta dt
\vec{A}(\vec{n}).d_t\vec{n}=\int\int\vec{n}(t,\tau).d\vec{a}=
\int_0^\beta dt \int_0^1 d\tau \vec{n}.\partial_t\vec{n}\times\vec{n}.
\label{5.2}
\end{eqnarray}
$n(t,\tau)$ is a parametrization of the surface on $S^2$ enclosed by
the trajectory $\vec{n}(t)$, that we chose so that
$\vec{n}(t,0)=\vec{n}(t)$ and $\vec{n}(t,1)=\check{k}$. We then see that
this term has a geometrical interpretation, which corresponds to the
oriented area on the sphere $S^2$ enclosed by $\vec{n}(t)$. 
Terms of this kind are known as Wess-Zumino terms.

Using the
explicit parametrization
\begin{eqnarray}
\lefteqn{\vec{n}(t,\tau)=}
\nonumber \\
& &  (\sin [(1-\tau)\theta(t)]\cos\phi(t), \sin
[(1-\tau)\theta(t)]\sin\phi(t), \cos[(1-\tau)\theta(t)])
\label{5.3}
\end{eqnarray}
where $t\in[0,\beta], \tau \in [0,1]$, the surface element is then given by
\begin{eqnarray}
\vec{n}.\partial_t\vec{n}\times\vec{n}=\dot{\phi(t)}\theta(t)
\sin[(1-\tau)\theta(t)].
\label{5.4}
\end{eqnarray}

The use of spherical coordinates for $\vec{n}$ allows for a straightforward
implementation of the constraint $\vec n^2=1$ and the functional measure is
\begin{equation}
{\cal D} \vec n(t)\delta(\vec n^2-1)={\cal D}\varphi(t){\cal
D}(cos\theta(t)) .
\label{5.6}
\end{equation}
Making use of the coordinates (\ref{5.3}), the partition function (17) can
be written as
\begin{equation}
{\cal Z}={\cal D}\varphi(t){\cal D}u(t)exp\left(\imath s\int_0^{\beta}dt
\dot\varphi(t)\left(1-u(t)\right)-sB\int_0^{\beta}u(t) dt\right),
\label{5.7}
\end{equation}
where $u(t)=cos(\theta(t))$, the integral over $\tau$ has been carried out
in the WZ term and the functional integration is restricted to continuous
closed trajectories. The corresponding boundary conditions are
\begin{equation}
\varphi(\beta)=\varphi(0)+2k\pi ,~~~~~~~~~~ \theta(\beta)=\theta(0)+2k'\pi
\label{5.8}
\end{equation}
and

\begin{equation}
\varphi(\beta)=\varphi(0)+(2\tilde k+1)\pi ,
~~~~~~~~~~\theta(\beta)=-\theta(0)+2\tilde k'\pi.
\label{5.9}
\end{equation}

We will first integrate out $\varphi(t)$. It is convenient to integrate by
parts the first term in the exponent of (\ref{5.7}). Then it reads
\begin{equation}
\imath s\left(\varphi(\beta)-\varphi(0)\right)(1-u(0))
+\imath s\int_0^{\beta}\varphi(t)~ \dot
u(t)dt.
\label{5.10}
\end{equation}
The change of $\varphi$ along the trajectory,
($\Delta\varphi=\varphi(\beta)-\varphi(0)$),
is a constant $2k\pi$ and $(2\tilde k+1)\pi$ for trajectories satisfying
(\ref{5.8}) and (\ref{5.9}) respectively.

The functional integral over $\varphi$ is now straightforward and the
result is

\begin{equation}
\int {\cal D}\varphi(t)exp\left(is\int_0^{\beta}dt~
\dot\varphi(t)\left(1-u(t)\right)\right)=
\sum_{\Delta \varphi}\delta[\dot u(t)]exp\left(\imath
s\Delta \varphi(1-u_0)\right).
\label{5.10'}
\end{equation}

The $\delta$-functional restricts the integral over $u(t)$ to constant trajectories. For this reason any
trajectory satisfying the boundary conditions (\ref{5.9}) is wiped out from the partition function.
The functional integral over the remaining trajectories can be written as a sum of ordinary
integrals, one for each posible value of $\Delta \varphi$.

\begin{eqnarray}
\lefteqn{{\cal Z}=\sum_k\int_{-1}^1 du_0 exp\left(is2k\pi (1-u_0)-sB\beta u_0\right)=}
\nonumber \\
& & \int_{-1}^1du_0\sum_{-\infty}^{\infty} \delta(s(1-u_0)-n)exp(-sB\beta_0).
\label{5.11}
\end{eqnarray}
The second equality follows from the Fourier representation of the $\delta$-function.

Integrating now over $u_0$ gives
\begin{equation}
{\cal Z}=\sum_0^{2s} exp(-\beta B(s-r)),
\label{5.12}
\end{equation}
which can be rewritten as
\begin{equation}
{\cal Z}=\sum_{m=-s}^{s} exp(-\beta Bm).
\label{5.13}
\end{equation}

This expression for the partition function coincides with the well-known result
obtained in elementary quantum statistical mechanics.

It should be stressed that our result was obtained without the use of any approximation
within the path-integral approach. One sould recall at this point that one of the
mathematical subtleties involved in the formal construction of path-integrals
is the restriction to continuous trajectories.
Our computation, which explicitly uses this restriction, provides a check that
this procedure leads to the correct result.

It is important to point out that at the classical level there is no
reason to choose the parameter $s$ to be integer or half-integer.
In this section we have found that this condition emerges naturally when 
the path integral for a 1-site problem is solved.

\section{Summary}

In this paper we have presented a constrainted Hamiltonian theory for spin systems.
Using the Dirac theory we have shown that our classical system has DB which can be
identified with the commutation relation for the components of the spin.
Following the recipe for quantizing constrained theories we have obtained a path
integral representation for the spin system, which coincides with the expression
obtained using coherent states \cite{fradkin}.

Our approach shows in a clear way that the path integral for the spin
systems is free of any approximation.

Finally, we have also recovered the partition function for a single spin
in a magnetic field from
our path integral expression.

Natural extensions of the present work would be to study the case of 
$Osp(2,2)$ (related to the t-J model)
and $SU(3)$ groups. In fact, we are working actually on this problem, but so
far we have found that a generalization of
this work is not straightforward and deserves a separate treatment.

\vspace{1cm}

{\it Acknowledgments:}

\noindent The authors acknowledge Fundaci\'on Antorchas for finantial
support. A.G. and A.D. would like to thank C. Abecasis, A.Foussats
and O.Zandron for
valuable discussions. D.C. thanks F. Schaposnik for a careful reading of the manuscript.

\end{document}